\documentclass[a4paper,fleqn,manuscript]{cas-sc}
\usepackage[sort,numbers]{natbib}
\usepackage{comment}
\usepackage{todonotes}
\usepackage{xcolor}
\usepackage{array}
\usepackage{soul}
\usepackage{url}
\usepackage{graphicx} 
\usepackage{caption} 
\usepackage{multirow}
\usepackage{graphicx,subcaption,lipsum}
\usepackage[pagewise]{lineno}
\usepackage{colortbl}
\usepackage{makecell}
\usepackage{booktabs} 
\usepackage[all]{nowidow}

\definecolor{darkgreen}{RGB}{0,100,0}
\definecolor{red}{RGB}{200,0,0}
\definecolor{lowpriority}{RGB}{144, 238, 144} 
\definecolor{mediumpriority}{RGB}{255, 192, 128}
\definecolor{highpriority}{RGB}{255, 153, 153} 

\def\tsc#1{\csdef{#1}{\textsc{\lowercase{#1}}\xspace}}
\tsc{WGM}
\tsc{QE}
\tsc{EP}
\tsc{PMS}
\tsc{BEC}
\tsc{DE}

\newcommand{\virgo}[1]{``#1''}

\newcommand{\customcells}[1]{\makecell*[{{p{11cm}}}]{#1}}
\newcommand{\customcelld}[1]{\makecell*[{{p{9cm}}}]{#1}}
\newcommand{\customcellc}[1]{\makecell*[{{p{4cm}}}]{\textbf{#1}}}

\ExplSyntaxOn
\cs_gset:Npn \__first_footerline: {Fornari\ F.,\ et\ al.:\ \textit{Preprint}\ \textit{version}}
\ExplSyntaxOff

\let\printorcid\relax

\begin{document}

\let\WriteBookmarks\relax
\def\floatpagepagefraction{1}
\def\textpagefraction{.001}

\shorttitle{Digital Twins of Business Processes: A Research Manifesto}
\shortauthors{Fornari F., et al.}

\title[mode = title]{Digital Twins of Business Processes: A Research Manifesto}

\author[1]{Fabrizio Fornari}
\cormark[1]
\ead{fabrizio.fornari@unicam.it}

\author[2]{Ivan Compagnucci}
\ead{ivan.compagnucci@gssi.it}

\author[1]{Massimo {Callisto De Donato}}
\ead{massimo.callisto@unicam.it}

\author[10]{Yannis Bertrand}
\ead{yannis.bertrand@kuleuven.be}  

\author[6]{Harry H. Beyel}
\ead{beyel@pads.rwth-aachen.de}

\author[7,8]{Emilio Carrión}
\ead{emcaryp@gmail.com}

\author[3]{Marco Franceschetti}
\ead{marco.franceschetti@unisg.ch}

\author[14]{Wolfgang Groher}
\ead{wolfgang.groher@ost.ch}

\author[13,14]{Joscha Grüger}
\ead{joscha.grueger@dfki.de}  

\author[9]{Emre Kilic}
\ead{Emre2.kilic@tu-dortmund.de}  

\author[4]{Agnes Koschmider}
\ead{agnes.koschmider@uni-bayreuth.de}

\author[11]{Francesco Leotta}
\ead{leotta@diag.uniroma1.it}  

\author[5,6]{Chiao-Yun Li}
\ead{chiaoyun.li@pads.rwth-aachen.de}

\author[12]{Giovanni Lugaresi}
\ead{giovanni.lugaresi@kuleuven.be}  

\author[13,14]{Lukas Malburg}
\ead{lukas.malburg@dfki.de}

\author[15]{Juergen Mangler}
\ead{juergen.mangler@tum.de }

\author[11]{Massimo Mecella}
\ead{mecella@diag.uniroma1.it}  

\author[8]{Oscar Pastor}
\ead{opastor@dsic.upv.es} 

\author[14]{Uwe Riss}
\ead{uwe.riss@ost.ch}

\author[3]{Ronny Seiger}
\ead{ronny.seiger@unisg.ch}

\author[10]{Estefania Serral}
\ead{estefania.serralasensio@kuleuven.be}  

\author[8]{Victoria Torres}
\ead{vtorres@pros.upv.es}  

\author[8]{Pedro Valderas}
\ead{pvalderas@pros.upv.es}  

    \affiliation[1]{organization={University of Camerino, School of Science and Technology, Computer Science Department},
    addressline={Via Madonna delle Carceri 7}, 
    city={Camerino},
    postcode={62032}, 
    country={Italy},}
    
     \affiliation[2]{organization={Gran Sasso Science Institute, Computer Science Department},
    addressline={Viale Francesco Crispi 7}, 
    city={L'Aquila},
    postcode={67100}, 
    country={Italy}}
    
    \affiliation[3]{organization={University of St.Gallen, Institute of Computer Science},
    addressline={Rosenbergstrasse 30}, 
    city={St.Gallen},
    postcode={9000}, 
    country={Switzerland}}
    
    \affiliation[4]{organization={University of Bayreuth, Business Informatics and Process Analytics},
    addressline={Wittelsbacherring 8}, 
    city={Bayreuth},
    postcode={95444}, 
    country={Germany}}
    
    \affiliation[5]{organization={Fraunhofer Institute for Applied Information Technology},
    addressline={Schloss Birlinghoven}, 
    city={Sankt Augustin},
    postcode={53757}, 
    country={Germany}}
    
    \affiliation[6]{organization={Process and Data Science (PADS), RWTH Aachen University},
    addressline={Ahornstraße 55}, 
    city={Aachen},
    postcode={52074}, 
    country={Germany}}
    
    \affiliation[7]{organization={Mercadona Tech, Mercadona},
    addressline={Plaza de Amèrica 2}, 
    city={Valencia},
    postcode={46004}, 
    country={Spain}}

    \affiliation[8]{organization={PROS Research Center - VRAIN Institute, Universitat Politècnica de València},
    addressline={Camino de Vera}, 
    city={Valencia},
    postcode={46022}, 
    country={Spain}}
    
    \affiliation[9]{organization={TU Dortmund University},
    addressline={Leonhard-Euler-Str. 5}, 
    city={Dortmund},
    postcode={44227}, 
    country={Germany}}
    
    \affiliation[10]{organization={Research Center for Information Systems Engineering (LIRIS), KU Leuven},
    addressline={Warmoesberg 26}, 
    city={Brussels},
    postcode={1000}, 
    country={Belgium}}
    
    \affiliation[11]{organization={Sapienza Università di Roma, DIAG},
    addressline={via Ariosto 25}, 
    city={Rome},
    postcode={00185}, 
    country={Italy}}
    
    \affiliation[12]{organization={KU Leuven, Faculty of Engineering Technology},
    addressline={Celestijnenlaan 300}, 
    city={Leuven},
    postcode={3001}, 
    country={Belgium}}
    
    \affiliation[13]{organization={German Research Center for Artificial Intelligence (DFKI), Branch Trier University},
    addressline={Behringstraße 21}, 
    city={Trier},
    postcode={54296}, 
    country={Germany}}

    \affiliation[14]{organization={Artificial Intelligence and Intelligent Information Systems, Trier University},
    addressline={Universitätsring 15}, 
    city={Trier},
    postcode={54296}, 
    country={Germany}}
    
    \affiliation[12]{organization={OST Eastern Switzerland University of Applied Sciences},
    addressline={Rosenbergstrasse 59}, 
    city={St. Gallen},
    postcode={9001}, 
    country={Switzerland}}
    
    \affiliation[15]{organization={TUM School of Computation, Information and
Technology},
    addressline={Boltzmannstraße 3}, 
    city={Garching bei München},
    postcode={85748}, 
    country={Germany}}

\begin{abstract}
Modern organizations necessitate continuous business processes improvement to maintain efficiency, adaptability, and competitiveness. 
In the last few years, 
the Internet of Things, via the deployment of sensors and actuators, has heavily been adopted in organizational and industrial settings to monitor and automatize physical processes influencing and enhancing how people and organizations work. 
Such advancements are now pushed forward by the rise of
the Digital Twin paradigm 
applied to organizational processes.  

Advanced ways of managing and maintaining business processes come within reach as there is a Digital Twin of a business process - a virtual replica with real-time capabilities of a real process occurring in an organization. 
Combining business process models with real-time data and simulation capabilities promises to provide a new way to guide day-to-day organization activities. 
However, integrating Digital Twins and business processes is a non-trivial task, presenting numerous challenges and ambiguities. This manifesto paper aims to contribute to the current state of the art by clarifying the relationship between business processes and Digital Twins, identifying ongoing research and open challenges, thereby shedding light on and driving future exploration of this innovative interplay. 
\end{abstract}


\maketitle

\section{Introduction}	

In the contemporary landscape, organizations constantly strive to enhance and sustain the efficiency and performance of their operational processes. This necessity is fueled by several factors, including the increasing competitiveness of the global market, environmental shifts,
variations in resource availability, emergent business opportunities, and the rapid evolution of digital technologies. To meet these challenges effectively, organizations must adopt strategies to constantly optimize their processes, reduce operational costs, and elevate the quality of their products or services \cite{DumasFundamentals18,WeskeBook}. 

The advent of the Internet of Things (IoT) has led to the emergence of a vast amount of interconnected and embedded computing devices with sensing and actuating capabilities, which are transforming how organizations handle their Business Processes (BPs) \cite{BP-meet-IoT-Manifesto}. While IoT actuators are utilized for process task automation, sensors are employed to gather real-time data, which can facilitate unprecedented opportunities for the discovery and exploitation of new process insights. These insights can be leveraged for the timely detection of anomalies in process execution, efficient process prediction, or to identify potential process improvements. The integration of IoT and BPs has led to
the emergence of the term \textit{IoT-Enhanced Business Processes} (or \textit{IoT-Aware Business Processes}) \cite{CompagnucciSLR,DeLuzi24,BP-meet-IoT-Manifesto,serral2024modeling,TorresVictoria_EstefaniaSerral_2020,ValderasIoT}. 
Such a concept is now pushed forward by the rise of \textit{Digital Twins}, which have become more affordable and promise to drive the future of BPs \cite{Schmitt2023}. 

\virgo{A Digital Twin (DT) is a virtual representation of real-world entities or processes, synchronized at a specified frequency and fidelity} \cite{digitaltwinconsortium}. The DT enables bidirectional data flow between the virtual representation and its physical counterpart
allowing for monitoring operations, simulation of what-if scenarios and prediction of future states and outcomes. Information gathered from DTs can be delivered to operators, to control systems, or directly to physical devices that can operate on the entity or process adjusting to optimizing performance \cite{AalstDT}. 

Initially adopted in the manufacturing sector to virtually replicate, simulate, and make predictions on physical assets \cite{Grieves2023}, the concept of DT is now being applied to organizational processes. 
This provides a new approach to rethinking and re-engineering modern BPs, leading to the possible creation of virtual replicas of actual BPs. These replicas allow for identifying potential issues, predicting outcomes, and optimizing operations through what-if analysis, extending the DT concept across entire BPs \cite{DumasDTBP}. 
This novel paradigm allows for a dynamic representation of organizational processes and ensures timely responses to unexpected events, optimizing development times and product quality. Additionally, the use of DTs at the organizational level might provide a holistic understanding of organizational dynamics, leading to more effective interventions and efficient organizational design.  

Integrating DTs and BPs is, however, a non-trivial task presenting numerous challenges. In light of this, an interactive session involving participants of two workshops, namely BP-Meet-IoT\footnote{\url{https://bp-meet-iot.webs.upv.es/}} and DT4BP\footnote{\url{https://pros.unicam.it/dt4bp2023/}}, was held in Utrecht during the 21st International Conference on Business Process Management. The interactive session was aimed to bring together researchers and practitioners to discuss the integration of BPs
and DTs. The discussion was guided by a set of open research questions that participants were asked to answer. A revised version of these questions is reported below.

\begin{enumerate}
    \item What is a Digital Twin? 
    \item What is the role of Digital Twins with respect to Business Processes?
    \item  What are the challenges of constructing a Digital Twin of a Business Process?
\end{enumerate}

During such an interactive session, participants discussed and contextualized ideas, viewpoints, and current findings. This collaborative setting fostered knowledge exchange between participants who provided different perspectives on the topics. Despite some doubts have been expressed on the DT concept, from the discussion it emerged that the integration of DTs and BPs presents interesting opportunities but significant challenges \cite{Biggest-BPM-Problems,DumasDTBP}. Especially participants emphasized the need for more clarity on the role of DTs with respect to BPs. 

To shed light on the integration between BPs and DTs, a group of twenty-three academics, ranging from experienced professors to young researchers from different academic and research institutions, was brought together. This paper is the result of such a joint effort. With this contribution we aim at providing a common view on the integration of BPs and DTs, contributing to the current state of the art by identifying in-progress research, opportunities, and open challenges that could guide further research. 

The paper is organized as follows. Section \ref{sec:background} presents background knowledge about DTs and BPs. Section \ref{sec:bpmeetdt} offers a comprehensive overview of the integration between BPs and DTs. Section \ref{sec:bpmeetdtchallenges} reports a list of challenges and research directions that we derived both from the literature and the acquired experience of the authors. Finally, Section \ref{sec:concluding} summarizes and concludes the paper. 
\section{Background}
\label{sec:background}
In this section, we report background information on the foundational concepts related to DTs and BPs.

\subsection{From Digital Models to Digital Twins}	
\label{sec:DT}

\begin{figure}[!htb]
	\centering
	\includegraphics[width=0.665\textwidth]{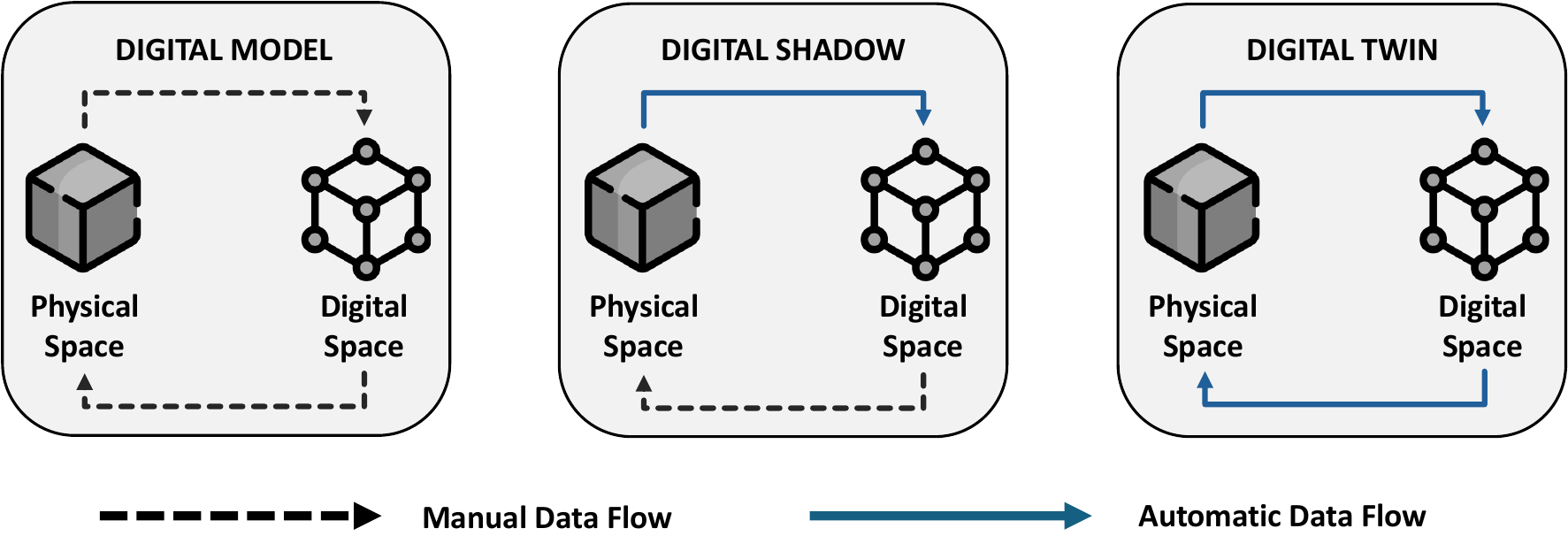}
	\caption{Digital Model, Digital Shadow and Digital Twin representation based on physical and digital spaces \cite{Grieves2017DigitalTM}.}
	\label{fig:dtconcepts}
\end{figure}

The evolution of model engineering has radically transformed how we design, develop, and manage complex systems \cite{BordeleauCEBW20}. From creating simple static diagrams, we have advanced to using increasingly sophisticated digital models capable of accurately representing the dynamic behavior of real-world entities \cite{Grieves2017DigitalTM,Tekinerdogan23}. In this context, concepts like \textit{Digital Model}, \textit{Digital Shadow}, and \textit{Digital Twin} have emerged \cite{,Grieves2017DigitalTM,MattaL23}, which we illustrate in Figure \ref{fig:dtconcepts}.

A \textbf{Digital Model} is a digital representation of one or more aspects of a physical entity
\cite{Grieves2017DigitalTM}. 
Examples of digital models include graphical 2D or 3D representations used by engineers and architects to visualize the spatial dimensions of machinery or buildings and their components. Another type of model involve the use of state machines, which depict the possible states and transitions that a physical entity can undergo. Additionally, workflow models represent the flow of steps and resources involved in tasks, specifying how each task is performed, who performs it, and the sequence of execution. Generally, a single model alone is not sufficient to fully describe a physical entity. Indeed, it may primarily target one perspective of the real entity instead of representing multiple perspectives (e.g., possible actions, current state, spatial dimension) \cite{TAO2022372}. 
However, despite their usefulness, digital models are often created manually, resulting in a static representation of reality. Human interventions are often necessary to update and maintain their accuracy and alignment with real-world conditions. This can limit the model's ability to faithfully represent the physical entity. This is represented in Figure \ref{fig:dtconcepts} by the dotted arrow between the physical and digital spaces, indicating a manual flow of data from the physical to the digital space.  
To address this limitation, integrating real-world data and information with designed digital models is fundamental. 

Nowadays, the use of IoT allows the real-time collection of data from the physical space, enabling the creation of what is known as a \textbf{Digital Shadow} (DS), which extends digital models of the physical entity
integrating real-time 
data \cite{Grieves2017DigitalTM}. 
A DS involves harvesting real-world data to create a digital representation that accurately mirrors a physical entity using digital models and dedicated tools \cite{BeckerShadow}. 
Continuously updating the model with such data establishes the conditions for faithfully replicating the entity's current status, enabling the possibility of conducting real-time monitoring of the physical entity. In Figure \ref{fig:dtconcepts}, this concept is illustrated with a solid arrow pointing from the physical space to the digital space, representing an automatic data flow that enables real-time synchronization of data from the physical object to its digital counterpart.

A \textbf{Digital Twin} (DT) is a digital representation that mirrors a real-world entity and that can be used to influence the physical counterpart. Such representation is synchronized with the physical data at a specified frequency and fidelity \cite{digitaltwinconsortium}.  
Differently from digital shadows that are mainly used for monitoring, a DT has additional enhanced functionalities to analyze and make informed decisions that can be reflected in the physical entity. 
In fact, DTs not only reflect the current state of their physical counterparts but can also be adopted to simulate the behavior of their physical counterparts, to support decision-making, and to control their influence in the real world.    
This bidirectional interaction is captured in Figure \ref{fig:dtconcepts}, where solid arrows point both from the physical space to the digital space and back, indicating a bidirectional automatic data flow.

\subsection{Business Processes}
\label{sec:BP}

Every organization, whether it is a government agency, a nonprofit organization, or a business, has to deal with the management of several processes \cite{DumasFundamentals18}. BPs are dynamic series of actions, events, and decisions leading to the creation of a product or the delivery of service \cite{DumasFundamentals18,WeskeBook}.  These processes are typically embedded within the operations of an organization and are often aligned with specific departments. For instance, within the production department, a BP might include the steps involved in assembling a product \cite{seiger2022integrating}. In the logistics department, a process could involve managing the supply chain, from receiving orders to coordinating deliveries to customers. 

\begin{figure}[!htb]
	\centering
	\includegraphics[width=0.8\textwidth]{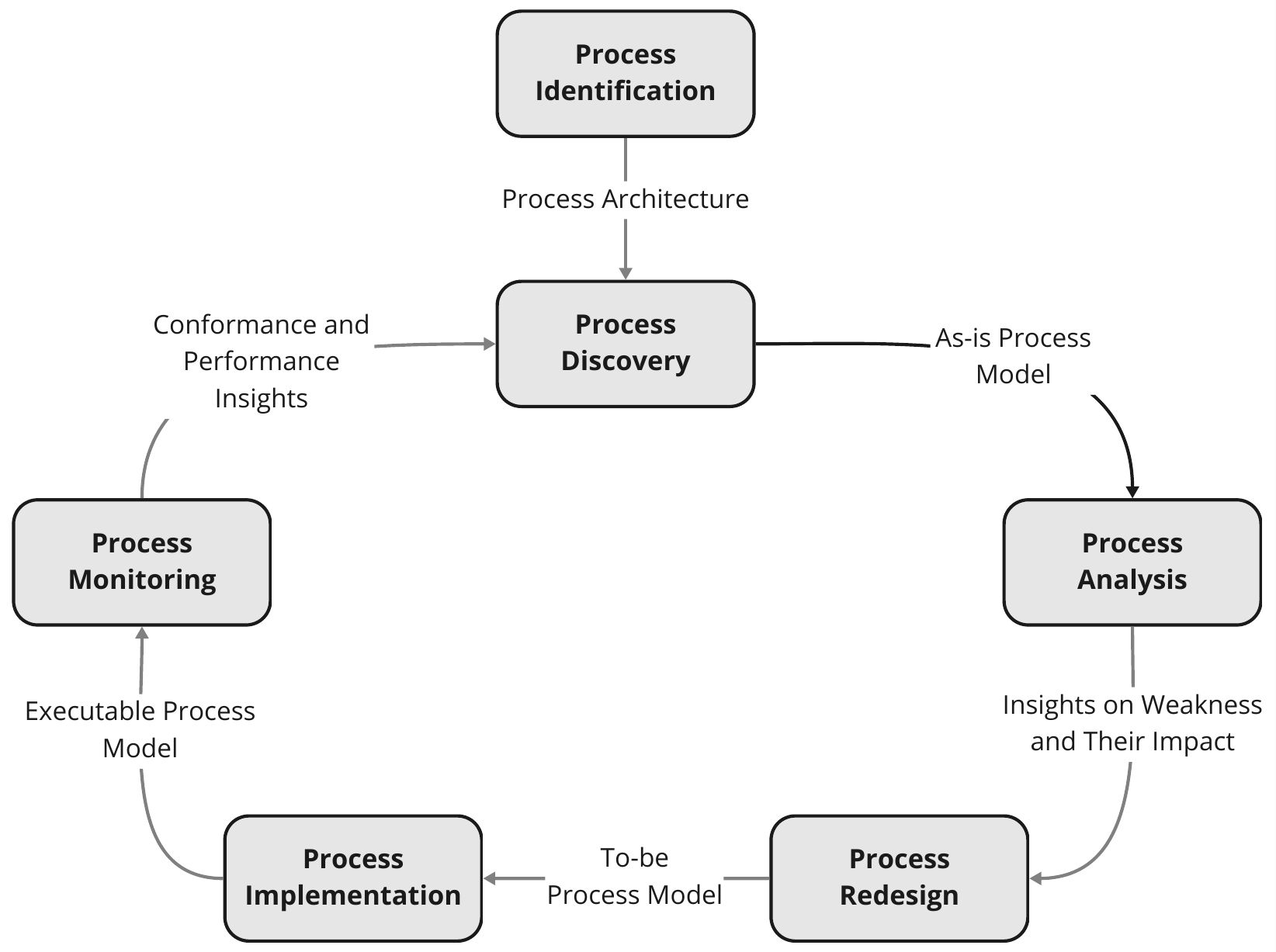}
	\caption{Business Process Lifecycle \cite{DumasFundamentals18}.}
	\label{fig:bpmlifecycle}
\end{figure}

Generally, BPs require continuous adjustment and improvement to remain competitive and efficient over time \cite{DumasFundamentals18,WeskeBook}.
The management of BPs involve various structured and sequential phases, collectively referred to as the BP Cycle \cite{DumasFundamentals18,WeskeBook}, as illustrated in Figure \ref{fig:bpmlifecycle}. 

Firstly, the \textit{Process Identification} phase involves identifying specific processes that require improvement, with careful attention to understanding existing operational challenges \cite{WeskeBook}. In the \textit{Process Discovery} phase, the actual BPs are documented and modeled, capturing the as-is process model that represents the current state of processes. Techniques such as process mining can be employed during this phase to extract process-related information from event logs, providing a data-driven approach to discovering process details \cite{DumasFundamentals18}. During the \textit{Process Analysis} phase, the \textit{as-is} process model is analyzed to identify weaknesses and their impacts, uncovering inefficiencies and bottlenecks \cite{AalstPM}. For instance, techniques like BP simulation can be used here to model potential changes and predict their effects on process performance \cite{AalstSimulation}. The \textit{Process Redesign} phase then leverages these insights to create the \textit{to-be} process model, aiming to enhance efficiency and align processes with organizational goals. In the \textit{Process Implementation} phase, the redesigned process model is executed, ensuring necessary resources, systems, and training are in place to support the new design \cite{DumasFundamentals18}. Technologies like the process engine facilitate this implementation by automating workflows and managing process execution \cite{traganos2024business}. Implementation occurs through careful planning and gradual execution, with clear communication and adequate training of participants \cite{DumasFundamentals18}. Finally, the \textit{Process Monitoring} phase involves a continuous assessment of the implemented processes' conformance and performance, gathering insights to evaluate effectiveness and identify further improvement areas. Tools such as dashboards \cite{eckerson2010performance} are instrumental in this phase, offering real-time monitoring and analytics to ensure processes are performing as expected. The execution of these phases can be
perpetuated to enhance the ongoing process and organizational growth \cite{WeskeBook}. 

Nowadays, the use of IoT has drastically reshaped the Information Technology sector, influencing the way organizations conduct business and becoming a key driver in the digital transformation of businesses \cite{BP-meet-IoT-Manifesto}. In the BP context, IoT devices are usually employed to automate and perform tasks required to achieve a BP goal \cite{BP-meet-IoT-Manifesto}. These kinds of processes are known as \textit{IoT-Enhanced Business Processes} \cite{CompagnucciSLR,DeLuzi24,BP-meet-IoT-Manifesto,serral2024modeling,TorresVictoria_EstefaniaSerral_2020,ValderasIoT}. 
They involve interconnected smart physical devices with sensing and actuating capabilities that contribute to improving operational efficiency, reducing costs, and enhancing the quality of products and services. 
The benefits of using IoT in BPs include increased automation, which reduces human errors, and comprehensive visibility into operations which supports more informed and timely decision-making \cite{BP-meet-IoT-Manifesto}. Additionally, analyzing data collected through IoT devices can provide additional valuable insights on the target process for improving further analysis and process optimizations.

Being the IoT referred to as the backbone of DTs \cite{iot-backbone-3,iot-backbone-1,iot-backbone-2}, the widespread adoption of IoT in BPs has enabled the construction of \textit{Digital Twins of Business Processes} (DTBPs) \cite{lyytinen2023digital}. This advancement allows for real-time simulation, monitoring, and analysis of entire BPs in a virtual environment. By leveraging DTs and IoT, businesses can conduct accurate simulations, predict outcomes, and optimize processes in the digital space before implementing changes in the real world. This, not only enhances decision-making and improves efficiency but also drives innovation, providing a significant competitive advantage in the market.

\section{Business Processes meet Digital Twins}
\label{sec:bpmeetdt}

\begin{figure}[htbp!]
  \centering
\includegraphics[width=1.0\textwidth]{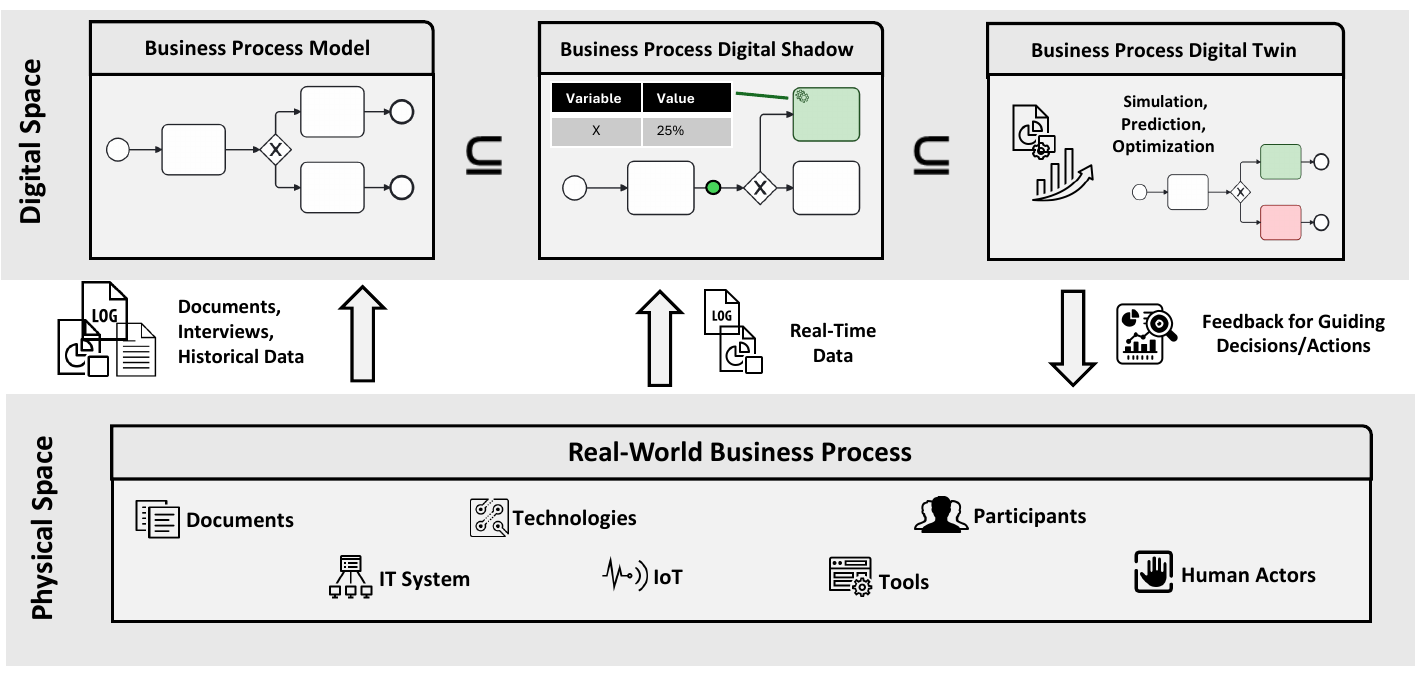}
  \caption{
  Connection and data integration between real-world BPs and their digital representations.
  } 
  \label{fig:process_dt_integration}
\end{figure}

In this section, we discuss the integration of BP and DT concepts. To clarify this integration, we report in Figure \ref{fig:process_dt_integration} a conceptual model that illustrates the relation among a \textit{Real-World Business Process}, its \textit{Process Model}, its \textit{Digital Shadow}, and its \textit{Digital Twin}.

BPs and DTs practices are not far apart. Both BPs and DTs involve the digital representation of real-life entities, gathering data from them, analyzing it, and conducting simulations and optimizations to foster continuous improvement and optimization when enacted in the real world. In the case of BPs, we talk about digital representations of processes or, in some cases, of entire organizations. That said, at the base of Figure \ref{fig:process_dt_integration} we reported a conceptualization of a \textbf{Real-World Business Process} that is made of entities of the physical space, such as people, systems, IoT devices, documents, etc. that contribute to the process execution. Note that a BP could also use DTs of Things (DTTs) which are DTs of physical assets such as machines, devices, etc. that are involved in the process \cite{lyytinen2023digital}.

The first step in studying and possibly improving an ongoing BP is to define a model that represents such a process. We refer to such model as a \textbf{Business Process Model} which corresponds to a digital model in the context of DT. 
Creating a BP model follows the standard Business Process Management (BPM) approach since BPs typically already exist within the organization \cite{DumasFundamentals18}. 
Mainly two kind of BP models are being adopted: imperative and declarative models. 
Imperative process models represent the whole process behavior at once and
can be modeled using different languages such as Petri Nets, BPMN diagrams, UML Activity diagrams, etc. Imperative process models represent well the behavior of processes in stable business-oriented environments. 
Declarative process models are based on the description of BPs through the use of constraints \cite{Declarative1}. Such constraints implicitly specify the allowed behavior of the process. Such models can be designed using Declare \cite{Declarative2}, DCR Graphs \cite{DCRGraph} and SCIFF \cite{Montali10}, etc. 
Declarative models are been regarded as appropriate for describing dynamic environments where processes are highly flexible and subject to change. While it is possible to start describing the model manually using existing information (e.g., documents, interviews, historical data), techniques such as process mining can facilitate the discovery of processes by analyzing process event logs \cite{AalstPM}. BP models can be manually inspected, used as a source of discussion among the process participants, and analyzed by applying standard BPM techniques to investigate their adherence to standard properties (e.g., soundness and safeness) or domain-specific properties \cite{Aalst97,BurattinMS16,CiccioM22,CorradiniFPRTV21}.

By linking a BP model with real-time data, we obtain what is known in the context of DTs as a digital shadow. A \textbf{Business Process Digital Shadow} (BPDS) represents the combination of the digital BP model and execution data of the process, offering a quantitative and real-time digital replica of the activities, operations, and workflows of a specific BP instance \cite{rabe2023framing}. 
A BPDS enables detailed, continuous monitoring of the current process execution. It allows for tracking and monitoring every stage of the BP in real-time, establishing a dynamic and updated view of the ongoing process instance \cite{rabe2023framing}. 
There can be various sources that provide BP-related data, such as IoT devices, business systems like Enterprise Resource Planning (ERP) or Customer Relationship Management (CRM) that provide transactional data and customer interactions, respectively \cite{lyytinen2023digital}. Additionally, data can be collected from machine logs, employee inputs, and external databases. This data can be integrated and visualized using dashboards or cockpits \cite{BanoMRVW22,Le24}. By continuously monitoring and integrating real-time data with the digital model, a BPDS 
consolidates an effective basis on which to perform analysis and decision-making, towards an optimization of the BP.

A full transition towards the DT concept becomes essential to identify inefficiencies in the execution of the BP, to predict outcomes possibly identifying problems and unexpected deviations from the process before they occur, and to provide data-driven insights for its optimization. 
A \textbf{Business Process Digital Twin\footnote{We adopt in this article the term \textit{Business Process Digital Twin} as an alternative way of referring to a \textit{Digital Twin of a Business Process}}} (BPDT)  aligns with the generic DTs concept, where models integrated with real-time data undergo continuous analysis for improvements impacting the real world process. BPDTs involve rigorous \textit{what-if} analysis through simulations and predictions \cite{DumasDTBP}. These methods are employed to evaluate different scenarios, predict outcomes accurately, and strategically optimize and control the BP. 
By evaluating Key Performance Indicators (KPIs) of different \textit{what-if} redesign processes, it is possible to identify areas for improvement and potential changes that are translated into actionable insights and interventions on the real-world process \cite{DumasDTBP}. 
It is worth noticing that the BPDT can initiate a feedback loop wherein new data generated by the optimized BP is collected and analyzed to further enhance the process, establishing a continuous cycle of optimization. This matches with the need to constantly maintain and update BPs to ensure that the real-world process remains efficient and effective over time \cite{DumasFundamentals18}.

\section{Open Challenges \& Research Directions}
\label{sec:bpmeetdtchallenges}

The concept of a BPDT brings a set of challenges that must be addressed to realize its full potential. This section reports on such challenges and points out research gaps that should be addressed in future research. We summarized challenges and future directions in Table \ref{tab:challenges}.

\textit{\textbf{C1-Design High-Fidelity Models.}} Gathering information about a process can be challenging and often requires multiple interactions with stakeholders, especially if the BP is less documented than the actual operations \cite{lyytinen2023digital}. Usually, in the context of DTs, there might exist documentation that defines all aspects into models. Modeling BPs typically involves the engagement of domain experts and encompasses not just specific tasks but also events, activities, participants, interactions, and decisions \cite{DumasFundamentals18}. This complexity makes it more difficult to verify the accuracy and fidelity of the models. The challenge lies in ensuring that the DT accurately replicates the BP, maintaining the integrity and reliability of the model by accurately representing all its components and interactions \cite{BocciarelliDP23,lyytinen2023digital,Valderas23BPDT}. This becomes challenging especially in the context of unstructured processes where the lack of predefined steps or standardized procedures could further generate ambiguity \cite{franceschetti2023characterisation} and difficulties in modeling \cite{BeckerDTO}.

In addressing the challenge of designing high-fidelity models for DTBPs, future research should focus on several key areas. Firstly, the development and enhancement of tools and methodologies for harvesting real-time data are crucial. Innovative approaches to data collection, particularly through IoT devices, should be explored to harvest contextual real-time information about the processes and participants involved~\cite{seiger2023interactive}. Another promising direction involves utilizing Large Language Models (LLMs) to extract process-relevant information from extensive organizational documents, thus bridging the gap between static documentation and dynamic process modeling \cite{KouraniB0A24}. Additionally, advancing event abstraction techniques will help managing the complexity of unstructured processes by providing clearer representations of activities and interactions. Recently, it has been suggested to treat activity recommendation as a knowledge graph completion task and to apply methods from this discipline \cite{SolaMAS21}. 

Leveraging process mining techniques can significantly aid in discovering and designing BP digital models that are more faithful to reality \cite{AalstPM}. We can rely on event logs of actual processes to discover the BP model of interest by applying available software tools \cite{LoyolaGonzalez23}. Recently a new paradigm in terms of event log has been introduced which focuses on recording the interactions between multiple objects within a BP. Such logs are referred as OCELs (Object-Centric Event Logs) \cite{berti2024ocel}. OCEL's multi-object perspective allows for identifying complex dependencies and interactions often missed by traditional event logs. 

In addition, DTs often rely on a combination of models to capture multiple perspectives of the physical entity \cite{CorradiniF0P022,VanDerHornM21}. With respect to BPs, a full transition towards a BPDT should rely on the combination of multiple models and logs that describe different aspects of the real BP ultimately leading to more precise and high-fidelity models. Therefore the integration of multiple models in BPs context should be the focus of future research.

\textit{\textbf{C2- Real-time Synchronization of Data and Models.}} 
DTs strongly rely on data gathered from the physical world to reflect the status and operations of the physical counterpart in the digital space. 
In the context of BPs, event logs might not be immediately available, leading to a temporal delay in the alignment of virtual models. 
IoT devices can be used to monitor process execution in real-time, providing timely and updated information \cite{BP-meet-IoT-Manifesto,Malburg.2023_MAPEK_JIIS,seiger2019consistency}. 
However, data can be distributed across different systems that use different sampling rates, different data formats, and different protocols to exchange data \cite{mangler2024internet}. 
Ensuring timely and accurate synchronization of data across these distributed and heterogeneous systems is crucial for maintaining the fidelity and relevance of the DT model \cite{Brockhoff,Comp23DT}. 

Addressing the challenge of synchronizing data and models in the context of DTs for BPs necessitates further research. One primary area of focus should be the automation of data pipelines to support process mining services \cite{Brockhoff}. This involves developing mechanisms for real-time data acquisition and integration from IoT devices, ensuring that the gathered data is timely and consistent despite the heterogeneity of sources and formats.  
One possibility to approach this is to leverage the strengths of Complex Event Processing (CEP) to handle the granularity and temporal aspects of data efficiently \cite{SofferCEP19,zerbato2021granularity}. 
Additionally, creating tools for synchronizing and visualizing multiple perspectives of BPs will be essential for maintaining a holistic and coherent view of the process states.

\textit{\textbf{C3-Ensuring Data Reliability.}} 
Data should faithfully represent reality. In a BP context, data collected from the real world is used for modeling and simulating BPs to obtain indications on possible interventions \cite{AalstSimulation,WeskeBook}.
On the one hand, if the recording of real process-executed activities is manually conducted, that might lead to erroneous or delayed activity registrations. On the other hand, IT systems that might be used to automatize such task can as well be subject to failure, especially in contexts where IoT devices are being adopted \cite{groher2023digital,seiger2019consistency}. Acknowledging the possibility of hardware or software failures that could lead to the generation of faulty data is necessary, and countermeasures should be taken \cite{groher2023digital,Riss20,Schultheis.2024_MissingSensorValues}. 

There is a need to focus on the system security of Process-Aware Information Systems (PAIS). On the one hand, conducting systematic reviews of how manual operations are conducted and possibly integrating Robotic Process Automation (RPA) can automate error-prone activities, reducing the likelihood of manual data entry errors \cite{BP-meet-IoT-Manifesto}. 
On the other hand, developing and adopting robust security frameworks will mitigate risks associated with hardware and software failures, ensuring the integrity of the collected data. The adoption of tamper-proof mechanism should be advocated such as the adoption of the blockchain technology \cite{blockchain1}. In the BP context the integration with blockchain technology has already been discussed \cite{blockchain2} and gathers its own challenges and research directions. 

\textit{\textbf{C4-Performing Valid Simulations.}} 
A DTBP incorporates the capability of simulating changes in process configurations and environment settings, allowing to estimate the effects of interventions in real BPs \cite{DumasDTBP}. 
This capability is essential for conducting what-if analysis, which aims at estimating the values that one or more process performance measures will take after a given BP change (i.e., adding a new task, changing the decision logic in a branching point, deploying new resources, etc.), allowing organizations to foresee the outcomes of interventions before real-world implementation \cite{Biggest-BPM-Problems}. 
Nevertheless, the execution of such simulations and the precise estimation of these alterations are subject to many complexities. BPs encompass various perspectives, including organizational structures, information systems, human interactions, and regulatory requirements, all of which require sophisticated modeling to distinguish between predictable and stochastic variations \cite{LopezPintadoDB24}. For actually conducting valuable simulations whose results can guide decisions affecting the actual process the available models must faithfully represent their real counterparts. 

Furthermore, simulating BP executions to discover optimal workflows and resource allocation might be a computationally demanding task requiring the development of efficient mechanisms to run timely up-front simulations and predictions. In addition, current approaches to process simulation often require a manual intervention to set up the parameters on which running the simulation and the results are manually assessed by analysts \cite{AalstSimulation,CamargoSM}. 

Research work should focus on integrating simulation models with real-time IoT data to create a dynamic and responsive simulation environment. 
By leveraging current methodologies \cite{CamargoSM,Leon23} that derive reliable simulation models from data, researchers can bridge the gap between static simulation setups and the dynamic nature of BPs. Developing efficient mechanisms to run timely up-front simulations and predictions will also be crucial, particularly for computationally demanding tasks. 
Improving data-driven simulation methods constitutes also a research direction. It involves incorporating automated discovery techniques for complex behaviors such as batching, prioritization, and resource availability patterns to address the limitations of traditional simulation models that often assume robotic resource behavior \cite{AalstSimulation,CamargoSM,DumasDTBP}. This includes developing advanced algorithms that can automatically generate key simulation parameters, such as optimal resource allocation, thereby reducing the need for manual intervention. 
Additionally, combining data-driven simulation with Machine Learning (ML) and Deep Learning (DL) techniques presents another promising direction. Recent work has shown that causal ML techniques 
can assess the impact of interventions on case outcomes \cite{DumasDTBP} and that DL model can be integrated at runtime during the simulation \cite{MeneghelloFG23}. 

\textit{\textbf{C5-Performing Process Predictions.}} 
DTs are used to  predict future behavior and performance of the physical entity \cite{Grieves2023,Grieves2017DigitalTM}. 
Such predictions aim at providing useful information to guide at design time the optimization of the physical entity before it is implemented. At run-time, it allows to predict adverse events, such as bottlenecks, that would occur if the entity continued on its current course and suggests taking remedial actions to prevent those issues. DTs are in fact adopted for performing predictive maintenance operations \cite{DinterTC22}. 
Leveraging the possibility to perform BP predictions is regarded has challenging given that we are not limiting our scope to a single physical object for which a DT is produced, like is often done in the manufacturing sector \cite{CiminoNF19,Grieves2023}, but a BP involves several entities, systems, human actors, and interactions among them raising the level of complexity in performing BP predictions. 

In the BP context, several approaches for conducting process predictions have been developed in recent years in the context of predictive and prescriptive process monitoring \cite{Brockhoff,Comuzzi2024,PPmonitoring2,PPmonitoring3,NeuLF22}. Those approaches often rely on so called \virgo{black-box} techniques such as the use of neural networks and this hinders the application of prediction models in real-life scenarios because decision-makers do not trust predictions provided as-is with no further explanation \cite{NeuLF22}. Therefore, an increasing attention has been paid to ensuring that the resulting predictive monitoring system is workable in practice by fostering explainability of the predictions \cite{Bozorgi2023,buliga2024guiding,buliga2024uncovering,GalantiLMNMSM23,rizzi2022explainable,VerenichDRN19}. 
However, the proposed techniques are still rare in practice \cite{WeinzierlRM19} and further research should be conducted on applying them and documenting successful cases.

\textit{\textbf{C6-Actuating the Intervention.}}
DTs allow for bidirectional communication between the digital and the physical space. In the BP context, the communication from the digital to the physical space is represented by the feedback resulting from the process optimizations \cite{Le24}. 
Such feedback needs to be promptly delivered to operators that conduct interventions, as well as other kinds of participants such as legacy applications, modern enterprise solutions, IoT devices, services, and the DTTs. 
Deploying an infrastructure able to process such automatized intervention and process execution is not a trivial task, considering that the various process participants might have different communication protocols or computational capabilities.

To effectively actuate interventions based on a BPDT, future research should focus on developing robust infrastructures that facilitate seamless bidirectional communication between digital and physical spaces \cite{lugaresi2023exploiting,Perno20}. This involves creating sophisticated communication mechanisms that can guide automatic control systems, enabling them to execute interventions in real-time accurately \cite{seiger2019toward}. 
It is therefore essential to adopt standardized communication protocols and enhance the computational capabilities of the various process participants, including legacy systems, modern enterprise solutions, IoT devices and other DT components. 
Moreover, research should explore mechanism to foster explainability of the proposed intervention. The use of advanced reporting tools, possibly leveraging LLMs, to generate comprehensive reports on the optimized versions of processes should be investigated. These reports could highlight the differences between the current and optimized processes, providing clear and actionable insights for business analysts. This will not only facilitate better understanding and acceptance of the changes but also enhance decision-making processes.  

\textit{\textbf{C7-Dealing with the Human Factor.}}
Human behavior exhibits a high degree of variability \cite{DiCiccioKIP,DumasDTBP,BP-meet-IoT-Manifesto}. The more a system incorporates human factors, the more complex it becomes to define DTs with a high level of fidelity due to the difficulty of representing the variability of humans in terms of experience, behavior, fatigue, and reactions to the situations, which can impact the process \cite{AalstSimulation,BP-meet-IoT-Manifesto,weber2023leveraging}. 
Modeling such processes is a challenging endeavor that a BPDT should be capable of, providing faithful simulation mechanisms that account for the human factor to properly simulate \cite{AalstSimulation,DumasDTBP}. 
The human factor not only affects the design of models and the simulation of processes, it also affects the enactment of process optimization, which can become challenging due to the variability in human behavior and the unpredictable nature of humans' reactions to interventions. Preserving the importance of humans in the loop in the enactment phase and balancing automation mechanisms expected from the DT becomes even more important, especially with the technological advancements that promise to automate most of the process activities \cite{AI-augmentedBPM,BP-meet-IoT-Manifesto}.

Future research should prioritize the development of advanced models that can accurately represent the variability in human behavior, experience, fatigue, and reactions \cite{DumasDTBP,weber2023leveraging}. In literature, researchers started to refer to the concept of Human Digital Twin \cite{Wang24} to faithfully represent aspects of a human actor. However, it is essential to delineate the extent to which DTs can realistically and ethically represent human aspects without oversimplifying or misinterpreting their contributions. Furthermore, research should focus on enhancing simulation mechanisms to account for human variability, ensuring that simulations reflect realistic human interactions and responses. This can be achieved by integrating behavioral data and leveraging machine learning algorithms to predict human actions under different scenarios \cite{DumasDTBP}. 
DTs must demonstrate resilience and avoid erroneous decisions when confronted with novel situations. Integrating human intelligence in the loop could be one way to foster resilience \cite{AalstDT}.

Additionally, ethical considerations must be at the forefront of this research. Ensuring the dignity and autonomy of human participants in BPs is paramount. Investigating frameworks that uphold human ethics while integrating DTs will help navigate the challenges associated with the human factor. By addressing these research directions, the integration of human elements into DTs can be significantly improved, leading to more robust, ethical, and effective BP models.

\textit{\textbf{C8-Dealing with New Situations.}} In the complex landscape of modern business, a company's success increasingly depends on its ability to respond quickly and flexibly to changes in its environment and the emergence of new situations \cite{Flexibility}. However, dealing with new situations can be particularly challenging \cite{DumasDTBP,BP-meet-IoT-Manifesto}. 
Often, the unpredictability arises because similar situations have not been observed before, leaving a gap in understanding how the process actors will respond \cite{Biggest-BPM-Problems,DiCiccioKIP}. In such cases, making accurate predictions becomes difficult \cite{DumasDTBP}. 
DTs are envisioned with the capability of being able to spot new situations and adjust their digital models to properly reflect the newly encountered real situation and the new real state. This capability is often referred to with the concept \textit{Living Model} \cite{allen2021digital}. How to reflect this living model concept into BPs is not yet fully clear, leaving open the question of whether to adjust a model with possible repair mechanisms \cite{armas2017interactive}, discover a model from scratch, or adopt a sort of continuous process mining approach  \cite{RizziFGM22}. 

In light of this, a possible solution is the implementation of continuous process mining to enable real-time data analysis. This approach involves constantly analyzing data streams to identify new patterns and anomalies as they emerge. By doing so, DTs can more accurately reflect new real-world situations, ensuring that models remain up-to-date and relevant. Additionally, research should focus on developing automated model adjustment and repair mechanisms. These mechanisms would allow for the dynamic adjustment of existing models in response to new data, maintaining their accuracy and usefulness without requiring a complete model overhaul. This aligns with the concept of the \textit{Living Model}, where DTs continuously evolve to mirror real-world changes. 
Exploring the balance between adjusting existing models, discovering new models from scratch, and adopting continuous process mining approaches is essential. 
Each of these strategies offers unique benefits and challenges, and their integration could lead to more robust and flexible BPs. Furthermore, ensuring an exit-free mechanism for handling faulty process execution is vital. This involves creating systems that can detect and respond to real-time process failures, preventing disruptions and maintaining operational continuity \cite{Malburg.2023_MAPEK_JIIS,seiger2019toward}.

\textit{\textbf{C9-Digital Twins as Process Participants.}}
DTTs offer clear advantages when they participate in BP activities \cite{GiacomoFLMS23,Valderas23BPDT}. They can be the target of requests to perform process actions as active participants. By replacing direct communication with the real counterpart, DTTs can provide more reliable data and information during process execution as well as enhance process simulations. Furthermore, linking DTs of process participants can create a comprehensive view of the overall process, enhancing the accuracy and utility of simulations. This integrated approach can provide deeper insights into the interactions and dependencies within BPs, facilitating more informed decision-making and process improvements. 

In the manufacturing context the ISO23247\footnote{\url{https://www.iso.org/standard/75066.html}} series provides a comprehensive DT framework, offering a robust foundation. 
To effectively incorporate DTTs as active participants in BPs, future research should focus on developing further standardized frameworks and protocols for their integration. 
Research should also focus on developing dedicated repositories for DTs, facilitating the efficient storage, management, and retrieval of DTs for specific assets. By establishing such repositories, DTs of the same entities can be reused in different contexts without the need for recreating them from scratch. This reuse capability enhances the role of DTs as reliable process participants, providing timely and accurate information during process execution and simulations.

\textit{\textbf{C10-Business Process Prototyping.}} 
In the context of DTTs, it is common to first create the digital model and then implement the physical asset based on the model, allowing for simulation, analysis, and performance optimization before physical implementation \cite{DonatoC00R23,Grieves2017DigitalTM}. This digital-first approach reduces the risks associated with creating new things by identifying and correcting errors early on and promotes rapid iteration, enabling faster and more flexible development cycles. 
In contrast to using process discovery techniques on already existing processes, which is the standard practice in BPM, it seems reasonable to evaluate the introduction of this prototyping practice for BPs. 
By utilizing virtual process design, digital tools can simulate BPs in a virtual environment before their actual implementation  \cite{lyytinen2023digital}. This enables detailed analysis and optimization, allowing businesses to explore various scenarios and configurations to determine the most effective process designs. 

The challenge lies in the nature of BPs and the complexity involved in handling human interactions, regulatory constraints, and varying operational contexts. A comprehensive and detailed representation of these aspects is necessary to effectively enable a digital-first approach in the case of BPs. Attempt to perform BP prototyping before actually implementing the BP should be investigated and evaluated in practice. Tool for virtual process design should be enriched to provide mechanism to simulate BPs in a virtual environment before their actual implementation, enabling detailed analysis and optimization. 

\textit{\textbf{C11-Software Tools for Digital Twins of Business Processes.}} 
DTs for BPs require a more integrated approach, capable of real-time data assimilation, continuous process optimization, and predictive analytics. Existing tools are not specifically designed with these advanced capabilities in mind, leading to limitations in their application to DTs \cite{TorresVictoria_EstefaniaSerral_2020}. 
Current implementations often rely on pre-existing BP modeling tools, such as Business Process Model and Notation (BPMN) \cite{Abouzid_Saidi_2023,Daclin_Daclin_Rabah_Zacharewicz_2023} or comprehensive software suites like Bizagi modeler, BonitaSoft, CAMUNDA, Visual Paradigm, Process Simulator \cite{Abouzid_Saidi_2023} and BPM ELMA \cite{Dorrer_2020}. While these tools offer robust features for traditional BP management, they fall short in addressing the dynamic, real-time requirements that DTs need. 
Well-known process mining tools such as Apromore \cite{RosaRADMDG11}, Celonis, SAP Signavio, started to integrate the concept of DT to support simulation and improvement of business operations, but in different ways and a standardized approach is still needed. 
As a result, the lack of standardized frameworks and platforms that could serve as the basis for DTs of BPs forces researchers to start from scratch or rely on fragmented solutions that are tailored to the needs of DTs \cite{Perez_Wassick_Grossmann_2022}.

There is a pressing need for the development of specialized tools designed explicitly for DTs in BPs. These tools should incorporate real-time data processing, advanced simulation capabilities, and seamless integration with IoT devices and enterprise systems. Additionally, exploring the use of artificial intelligence and machine learning within these tools could enhance their ability to predict process outcomes and suggest optimizations automatically.

\textit{\textbf{C12-From Digital Twin of Business Processes to Digital Twin of Organizations.}} 
Despite applying the DT concept to BPs is a challenging task \cite{BeckerDTO,lyytinen2023digital,Riss20}, recently the concept of Digital Twins of Organization (DTOs) has also been envisioned. It consists in applying the concept of DT to an entire organization 
\cite{park2021realizing}. DTOs promise to enable the analysis, simulation, and optimization of organizational operations to facilitate strategic planning, enhance decision-making processes at the strategic level, and improve overall organizational efficiency. 
However, in the context of a DTO, rather than a conventional DTs or a BPDT, this complexity becomes even more pronounced. A DTO does not merely represent a single product or system but must incorporate and simulate a wide range of BPs, departmental interactions, workflows, and operational dynamics. Optimizations and redesign of BPs become even more critical at the organizational level as it is crucial to pay attention to the interconnection of these processes. Changes made to one process can impact other linked processes, potentially causing cascading errors if the new process does not integrate correctly with the others. Managing such an increasing level of complexity and identifying or defining supporting tools is a challenge. 

To advance DTOs, understanding organizational dynamics and stakeholder interactions is essential. Integrating insights from organizational science and fostering dialogue between technical and organizational design communities will enhance DTOs' effectiveness \cite{lyytinen2023digital}. Continuous model assessment and improvement are vital, as it is investigating the balance between technical feasibility and economic value in DTO development \cite{lyytinen2023digital, WurmDTO}.
In addition, researchers should explore DT's role in algorithmic management in traditional industries and assess platform business insights' applicability \cite{WurmDTO}. Finally, designing intuitive user interfaces and advanced analysis tools for DTOs is crucial for visualizing and optimizing organizational processes \cite{Riss20}. Establishing common standards for data representation and communication protocols will facilitate DTO interoperability and widespread adoption, enhancing process transparency and flexibility through AI-augmented data analysis \cite{groher2023digital,Riss20}.

\begin{table}[htbp!]
\caption{Challenges and research directions in the realization of the BPDT concept.}
\label{tab:challenges}
\resizebox{1\textwidth}{!}{
\begin{tabular}{lll}
\toprule
\textbf{Challenges}                       & \textbf{Descriptions}                       & \textbf{Research Directions}  \\ \hline
    \customcellc{\textbf{C1: Design High-Fidelity Models}} &
    \customcelld{BPs typically involve multiple participants and source of information. This complexity makes it more difficult to design and align BP models with reality.} &
    \customcells{Large Language Models (LLMs) can be adopted to extract process-relevant information from extensive organizational documents. IoT devices should be integrated to capture for real-time data about processes and participants.  Process Mining should be applied to automatize model discovery. }  \\ \hline
    \customcellc{C2: Real-time Synchronization of Data and Models} & 
    \customcelld{Process data and event logs may not be immediately available, resulting in a time delay in the alignment of virtual models. Data may be in different formats and distributed across different heterogeneous systems challenging retrieval operations.} &
    \customcells{Foster the integration of IoT devices to perform real-time data acquisition. Handle the granularity and temporal aspects of data efficiently. Investigate the automation of data pipelines to support process mining services.} 
    \\\hline
    \customcellc{C3: Ensuring Data Reliability} & 
    \customcelld{Manual recording of real process-executed activities may result in inaccurate data, whereas IT systems or IoT devices may be subject to failure leading to the recording of erroneous data.}  & 
    \customcells{Enhance security of Process-Aware Information Systems (PAIS). Apply Robotic Process Automation (RPA) to manual error-prone activities. Investigate and adopt tamper-proof mechanism such as blockchain.} \\\hline
    \customcellc{C4: Performing Valid Simulations} & 
    \customcelld{The different perspectives in the BPs make it difficult to accurately simulate interventions in the real world, while discovering optimal workflows can be a computationally demanding task.}& 
    \customcells{Investigate generative models to replicate multi-perspective in process's behavior. Investigate enhanced data-driven simulation methods to automatically generate optimal simulation parameters. Foster the use of real-time IoT data to enables more dynamic and responsive simulation environments.} \\\hline
    \customcellc{C5: Performing Process Predictions} & 
    \customcelld{Predictions of real-world BPs involve multiple entities to be considered, increasing the level of complexity. Predicted outcomes of black-box approaches are not trusted.} & 
    \customcells{Enriching predictive and prescriptive process monitoring to improve the integration of AI-based techniques and data-driven approaches with an attention on the explainability of resulting predictions and prescriptions.}\\\hline
    \customcellc{C6: Actuating the Intervention} & 
    \customcelld{The lack of a proper infrastructure and the adoption of different communication protocols with different computational capabilities hinders the automatizing of interventions. The is a lack of proper documentation to guide user in performing intervention.} & 
    \customcells{Deploy IoT devices and proper infrastructures to support BPs. Adopt standardized communication protocols to improve the streamline of the interventions. 
    Investigate LLMs to facilitate better understanding and acceptance of the to be actuated intervention.} \\\hline
    \customcellc{C7: Dealing with the Human Factor} & 
    \customcelld{Dealing with the variability of human behavior throughout the entire process lifecycle, makes it difficult to define a BPDT with a high level of fidelity.} & 
    \customcells{Investigate the concept of Human Digital Twins to capture the nuances of human actors within BPs. Apply and evaluate ML and DL algorithms to predict human actions under different scenarios. Empower the role of humans in the loop by investigating on the integrate of human intelligence for defining resilient DTs and by developing ethical frameworks. } \\\hline
    \customcellc{C8: Dealing with New Situations} & 
    \customcelld{Unforeseen situations in BPs make it difficult to understand their status and how they will evolve. This raises the question of whether to adjust an existing models or discover new ones through process mining.} & 
    \customcells{Investigate the application of continuous process mining to identify new patterns and anomalies as they emerge in process execution. Investigate the concept of living model w.r.t. BPDT together with mechanisms to perform model adjustments and repairs. 
    } \\\hline
    \customcellc{C9: Digital Twins as Process Participants} & 
    \customcelld{Considering DTs as participants in BPs is still an emerging topic and standardization approaches for their integration have not yet been fully defined.} & 
    \customcells{Define standardized frameworks and protocols to integrate DTs as BP participants. Use of DT repositories to facilitate the management and retrieval of DTs and promote their reusability within BPs.} \\\hline
    \customcellc{C10: Business Process Prototyping} & 
    \customcelld{Prototyping BPs is not a common practice and requires comprehensive and detailed representation of multiple BP perspectives to effectively enable a digital-first approach in the case of BPs.} & 
    \customcells{BP prototyping before actually implementing the BP should be investigated and evaluated in practice. Tool for virtual process design should be enriched to provide mechanism to simulate BPs in a virtual environment before their actual implementation, enabling detailed analysis and optimization.
    } \\ \hline
    \customcellc{C11: Software Tools for Digital Twins of business processes} & 
    \customcelld{Existing tools offer robust features for traditional BP management but they lack in addressing a fully integration of DT related concepts.} & 
    \customcells{Tools explicitly designed for DTs of BPs to support real-time data processing are needed. The integration of advanced AI-powered simulation capabilities should be investigated.} \\ \hline
    \customcellc{C12: From DT of business processes to DT of organizations} & 
    \customcelld{Organizations have to deal with multiple and interconnected BPs, departmental interactions, multiple and possibly shared resources, therefore realizing a DT of an organization inherits such a complexity.} & 
    \customcells{Investigate the systematic integration of DTs to at the organization level. Explore digital tools aided with AI to support advanced analysis for DTOs. Enforce data standardization and communication protocols to facilitate interoperability and transparency within the organization.} \\
\bottomrule
\end{tabular}
}
\end{table}

\section{Concluding Remarks}
\label{sec:concluding}

DTs provide many opportunities, and their combination with BPs promises to revolutionize the way we deal with them. 
With this work, we reported on the results of a joined effort of a group of twenty-three academics, ranging from experienced professors to young researchers from different academic and research institutions, enthusiastic about the technological advancements and perspectives that the integration of DT and BP concepts can lead to. 
We specifically provide a way to look at such integration, laying the foundation for a common vision on the topic. 
We clarified on the usage of the terms Business Process Model, Business Process Digital Shadow, and Business Process Digital Twin. To reach a full transition to a Business Process Digital Twin different challenges must be addressed. We identified twelve of such challenges that we summarized in Table \ref{tab:challenges} and we encourage researchers to address them by providing pointers to future research directions. 

Furthermore, while analyzing the literature for this work we noticed a significant difference in the maturity levels of DTs. DTs of things are highly advanced, with extensive research and practical applications demonstrating their efficacy \cite{SemeraroLPD21}. In contrast, DTs of BPs are less mature due to the complexity involved in integrating multiple stages of a process, from data collection to process mining and workflow mapping and despite research work on the topic is increasing \cite{LugaresiJS23,PEREZ20211755,rabe2022concept,rabe2023framing} actual applications and documented cases are required to demonstrate the applicability in practice. 
We want to encourage researchers in pursuing the realization of BPDTs. A lot of research efforts have already been made in the field of BP management, and several approaches that have been defined to deal with the digitization of BPs can find a place to contribute to the realization of this BPDT concept. All that is required is a proper combination, tuning and adoption in real BP scenarios.

Instead, DTs of organizations represents an even newer and more complex frontier, aiming to encompass all aspects of an organization. This requires the integration of diverse data sources, subsystems, and human factors, making this kind of DT the least mature among the three types and posing significant challenges for effective implementation \cite{BeckerDTO,groher2023digital,park2021realizing}. In addition, DTs can be applied further beyond the boundary of an organization, reaching for instance the concept of DT of whole cities \cite{deng2021systematic,DTsmartcities,DTsmartcities2}, making the DT application inevitably more challenging as the complexity of the target physical entity under consideration increases.

Finally, integration of multiple levels of DTs is crucial for achieving a seamless and effective digital representation of complex scenarios. This poses additional challenges on which to direct further research efforts.

\bibliographystyle{cas-model2-names}
\bibliography{biblio.bib}

\end{document}